\documentstyle[psfig]{mn2e}

\def\gs{\mathrel{\raise0.35ex\hbox{$\scriptstyle >$}\kern-0.6em
\lower0.40ex\hbox{{$\scriptstyle \sim$}}}}
\def\ls{\mathrel{\raise0.35ex\hbox{$\scriptstyle <$}\kern-0.6em
\lower0.40ex\hbox{{$\scriptstyle \sim$}}}}
\def\ls{\mathrel{\hbox{\rlap{\hbox{\lower4pt\hbox{$\sim$}}}\hbox{$<$}}}}
\def\gs{\mathrel{\hbox{\rlap{\hbox{\lower4pt\hbox{$\sim$}}}\hbox{$>$}}}}

\title[AGN fraction in galaxy clusters]
      {The drivers of AGN activity in galaxy clusters: AGN fraction as a function of 
mass and environment}
\author[K.\,A.\ Pimbblet ]
       {Kevin A.\ Pimbblet$^{1,2}$\thanks{email: Kevin.Pimbblet@monash.edu}, 
Stanislav S.\ Shabala$^{3}$,
Chris P.\ Haines$^{4}$,
\and Amelia Fraser-McKelvie$^{1,2}$, 
David J.\ E.\ Floyd$^{1,2}$
        \vspace*{1mm}\\
        $^{1}$School of Physics, Monash University, Clayton, Victoria 3800, Australia\\
$^{2}$Monash Centre for Astrophysics (MoCA), Monash University, Clayton, Victoria 3800, Australia \\
$^{3}$School of Mathematics \& Physics, University of Tasmania, Private Bag 37, Hobart, Tasmania 7001, Australia\\
$^{4}$Steward Observatory, University of Arizona, 933 North Cherry Avenue, Tucson, AZ 85721, USA\\
}
\date{\today}

\pagerange{000--000}

\begin{document}

\maketitle

\begin{abstract}
We present an analysis of optical spectroscopically-identified AGN down to a cluster
magnitude of $M^\star +1$ in a sample of 6 self-similar SDSS galaxy clusters at $z \sim 0.07$.
These clusters are specifically selected to lack significant substructure at bright limits
in their central regions so that we are largely able to eliminate the local action of
merging clusters on the frequency of AGN. We demonstrate that the AGN fraction
increases significantly from the cluster centre to $1.5 R_{\mathrm virial}$, but tails off at larger radii.
If only comparing the cluster core region to regions at  $\sim 2 R_{\mathrm virial}$, 
no significant variation would be found.
We compute the AGN fraction by mass and show that massive
galaxies ($\log({\mathrm stellar~mass}) > 10.7$) are host to a systematically higher fraction of AGN
than lower mass galaxies at all radii from the cluster centre. We attribute
this deficit of AGN in the cluster centre to the changing mix of galaxy types
with radius. 
We use the WHAN diagnostic to separate weak AGN from `retired' galaxies in 
which the main ionization mechanism comes from old stellar populations. 
These retired AGN are found at all radii, while the mass effect is much 
more pronounced: we find that massive galaxies are more likely to be in 
the retired class.
Further, we show that our AGN have no special position inside galaxy clusters -- they
are neither preferentially located in the infall regions, nor situated at local maxima
of galaxy density as measured with $\Sigma_5$. However, we find that the most powerful AGN (with 
[OIII] equivalent widths $<-10$~\AA) reside at significant velocity offsets in the cluster, 
and this brings our analysis into agreement with previous work on X-ray selected AGN.
Our results suggest that if interactions with other
galaxies are responsible for triggering AGN activity, the time-lag between trigger and
AGN enhancement must be sufficiently long to obfuscate the encounter site and wipe out
the local galaxy density signal.

\end{abstract}

\begin{keywords}
galaxies: active ---
galaxies: evolution ---
galaxies: clusters: general
\end{keywords}

\section{Introduction}
Active galactic nuclei (AGN) are typically found inside massive galaxies that
exhibit significant, on-going or recent, star-formation (Kauffmann et al.\ 2003; 
Jahnke et al.\ 2004; 
Heckman et al.\ 2005;
von der Linden et al.\ 2010;
Floyd et al.\ 2012).
The power source for AGN is expected to be gas accretion on to a massive
black hole (Lynden-Bell 1969) which suggests that black hole and galaxy 
spheriodal growth are closely linked (cf.\ Richstone et al. 1998; 
Kauffmann et al.\ 2003).

\begin{figure*}
\centerline{\psfig{file=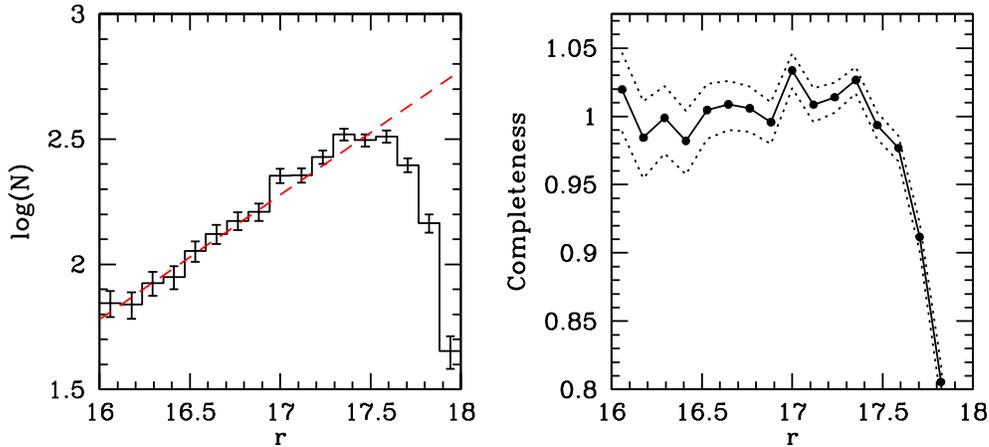,angle=0,width=5.45in}}
\vspace*{-7cm}
\caption{A histogram of r-band magnitudes (with Poisson errors)
for our sample is displayed in the left-hand panel.
A line of best fit (dashed line) is fitted to the linearly increasing
region of this plot (i.e.\ $16.0<r<17.0$) which is used to create a completeness
diagnostic plot (right-hand panel).  The points in the right hand panel
are the ratio of $\log(N)$ in the left-hand panel to line of best fit.  The dotted lines
enclosing these points denote the 1$\sigma$ error of the completeness values.
At our adopted limiting magnitude of $r=17.77$, the spectroscopy is still $>90$
per cent complete.
}
        \label{fig:completeness}
\end{figure*}

Mergers have frequently been cited as a method to fuel AGN (e.g.\ Sanders et al.\ 1988)
and a number of morphological studies claim an excess of post-merger systems in
their AGN samples (Bahcall et al.\ 1997; Canalizo \& Stockton 2001; 
Urrutia et al.\ 2008; Letawe et al.\ 2010; Smirnova et al.\ 2010).
Given that the fuel source for AGN is in the gas phase, 
any physical mechanism that has the potential to disturb the morphology of a 
galaxy such as harassment (Moore et al.\ 1996) may also 
produce an enhancement of AGN activity -- i.e.\ not simply mergers.  
Since such physical mechanisms can be tied to environment,
AGN may therefore be thought of as 
sign-posts to galaxy evolution in some circumstances (cf.\ Reichard et al.\ 2009). 
Indeed, there is extensive literature supporting the idea that 
AGN (defined in various ways using different wavelengths) 
are influenced by environment.
For instance, 
Kauffmann et al.\ (2004) report that the fraction of optical spectroscopic AGN is markedly
different for galaxies in different density regimes (explicitly: the AGN
fraction decreases as a function of increasing local galaxy density; 
see also Montero-Dorta et al.\ 2009).
This is supported by a study of 51 galaxy clusters by Ruderman \& Ebeling (2005)
who find an excess of X-ray point sources within 3.5 Mpc of the centre of
galaxy clusters in comparison to control samples which they attribute
to AGN that have been triggered by close encounters with neighbouring galaxies.
This broadly supports an increase of AGN fraction with increasing galaxy density, but
they divide this excess into two regions: near the cluster core where
galaxies are interacting with the brightest cluster member, and at around the virial
radius whose excess they report is attributable to low energy collisions 
at the cluster-field boundary.
Further, Popesso \& Biviano (2006) detail an anti-correlation between cluster AGN
fraction and cluster velocity dispersion (see also Sivakoff et al.\ 2008).  
They contend that this anti-correlation
indicates that the merger rate of clusters affects the AGN fraction since AGN
are likely to have played a strong hand in heating the intra-cluster medium
and thereby 
drive evolution in sub-groups that eventually form clusters (cf.\ Mamon 1992).
From an investigation of the Abell~901/902 system, Gilmour et al.\ (2007) show that 
there is a deficit of AGN in the highest density regions that supports the above works
and other investigations of cluster AGN locations (see 
Gisler 1978;
Dressler, Thompson \& Shectman 1985; 
Coldwell, Mart{\'{\i}}nez \& Lambas 2002; 
Georgakakis et al.\ 2008;
Gavazzi, Savorgnan \& Fumagalli 2011; Pimbblet \& Jensen 2012).
Naively, such a trend makes sense since galaxies in cluster centres would be more stripped of 
cold gas that can fuel an AGN than on cluster outskirts (see also Constantin et al.\ 2008;
Lietzen et al.\ 2011).

Equally, there is an increasing body of literature that indicates the opposite is true:
environment plays little or no role in the frequency of AGN.  
Examining X-ray emission from clusters, Miller et al.\ (2003) find no evidence
for an enhanced cluster AGN fraction.
This is supported by spectroscopic work on 8 clusters by 
Martini, Mulchaey \& Kelson (2007) who detail the AGN fraction is
no lower in cluster centres than a control field sample.
Although some of this may be caused by mass selection 
effects (cf.\ Pasquali et al.\ 2009; Pimbblet \& Jensen 2012), Haggard et al.\ (2010)
demonstrate that there is no significant difference in AGN fraction between cluster
and field samples for a constrained range of absolute magnitudes.
This is supported by von der Linden et al.\ (2010) who examine $>500$ Sloan Digital Sky Survey
(SDSS; Abazajian et al.\ 2009) clusters and find no trend in AGN fraction with distance from
cluster centres (see also Atlee et al.\ 2011; Klesman \& Sarajedini 2012).

In this work, we present a new analysis of the AGN fraction's dependence on 
environment and mass to elucidate the issues summarized above
using a sample of low redshift SDSS galaxy clusters that are
free from known structure contamination.  In Section~2 we detail the dataset that 
we use in this investigation.  In Section~3, we compute how the AGN fraction varies 
with radius from the cluster centre and galaxy mass before discussing and summarizing
our results in Section~4.
Throughout this work, we adopt
a standard, flat cosmology with
$\Omega_M = 0.238$, $\Omega_{\Lambda}=0.762$ and $H_0=73$~km~s$^{-1}$~Mpc$^{-1}$
(Spergel et al.\ 2007).

\section{Dataset}
Since mergers of galaxy clusters or sub-clusters may locally enhance AGN activity,
we require a sample of clusters that are relatively free from such activity.
In Pimbblet (2011), we presented a sample of 14 such SDSS galaxy clusters derived
from the earlier work of Plionis, Tovmassian 
\& Andernach (2009) that have no merging or significant 
interaction with other (comparable) structures 
within the limits of SDSS observations.
In brief, Pliois et al.\ (2009) use the criteria and catalogue of     
Andernach et al.\ (2005) to generate a "clean" sample of Abell et al.\ (1989)
clusters from.  
This consists of considering the velocity distribution of each cluster in
turn and its flatness (Struble \& Ftaclas 1994), and also removal of
any cluster with $>1$ X-ray peak.  We refer the reader to Plionis et al.\ (2009)
for a full description of this process.

Here, we restrict the Pimbblet (2011) sample to 6 galaxy clusters that are within
a narrow redshift slice ($0.070<z<0.084$; Table~\ref{tab:clusters}). 
The reason for selecting such a narrow sample to work with
is to create a composite stacked cluster whose variation in  
absolute magnitude that corresponds to a given apparent magnitude completeness
limit is small -- no more than $\Delta M_R = 0.4$ (Pimbblet 2011).
In terms of look-back time, the difference between our highest and lowest
redshift clusters is $\sim 0.1$ Gyr.

In-line with the
SDSS spectroscopic limit (see Abazajian et al. 2009; Strauss et al. 2002),
we use a limiting magnitude of $r=17.77$ in this work.
Fig.~\ref{fig:completeness} demonstrates that at 
$r=17.77$, SDSS spectroscopy is still $>90$ per cent complete 
for our cluster sample (cf.\ Pimbblet \& Jensen 2012;
Jensen \& Pimbblet 2012; Strauss et al.\ 2002).  Further, Pimbblet (2011)
notes that the sample covers little more than a factor of 2 in cluster
mass. Combined with the small difference in look-back time,
this ensures that the clusters in our sample are reasonably self-similar
and are broadly at a comparable evolutionary stage.

%
%
\begin{table*}
\begin{center}
\caption{The cluster sample used in this work.  
\hfil}
\begin{tabular}{lllllll}
\hline
Name  & RA      & Dec     & $\overline{cz}$ & $\sigma_{cz}$  & $R_{\mathrm virial}$ & N($<3R_{\mathrm virial}$)\\
      & (J2000) & (J2000) & (km\,s$^{-1}$)  & (km\,s$^{-1}$) & (Mpc)    &    \\
\hline
A1205           & 11 13 58.1 & $+$02 29 56 & 22506 & 938  & 1.88 &  110 \\
A1424           & 11 57 26.4 & $+$05 05 52 & 22764 & 780  & 1.56 &  78 \\
A1620           & 12 50 03.0 & $-$01 33 45 & 25275 & 1007 & 2.01 &  153 \\
A1650           & 12 58 34.7 & $-$01 43 15 & 25176 & 864  & 1.73 &  117 \\
A1767           & 13 36 31.6 & $+$59 08 51 & 21111 & 988  & 1.98 &  111 \\
A2670           & 23 54 13.7 & $-$10 25 09 & 22836 & 976  & 1.95 &  132 \\
\hline
\vspace*{-1.0cm}
\end{tabular}
  \label{tab:clusters}
\end{center}
\end{table*}

In Table~\ref{tab:clusters}, we give the global properties of the clusters used in this
work, including mean recession velocity ($\overline{cz}$), cluster velocity dispersion
($\sigma_{cz}$) and virial radius ($R_{\mathrm virial}$).  The former two are based on Miller et al.\ (2005)
whilst the virial radius is computed from $\sigma_{cz}$ using the relation presented
by Girardi et al.\ (1998). Cluster membership is then
simplistically defined to be all galaxies within $\pm 3 \sigma_{cz}$
of $\overline{cz}$. The number of galaxies in each cluster within $<3R_{\mathrm virial}$ is reported
as N($<3R_{\mathrm virial}$) in Table~\ref{tab:clusters}.  These numbers are approximately the
same as those reported by Pimbblet et al.\ (2006; see their Table~2) 
for rich, X-ray luminous clusters at $z\sim 0.1$.

\begin{figure}
\hspace*{2cm} \centerline{\psfig{file=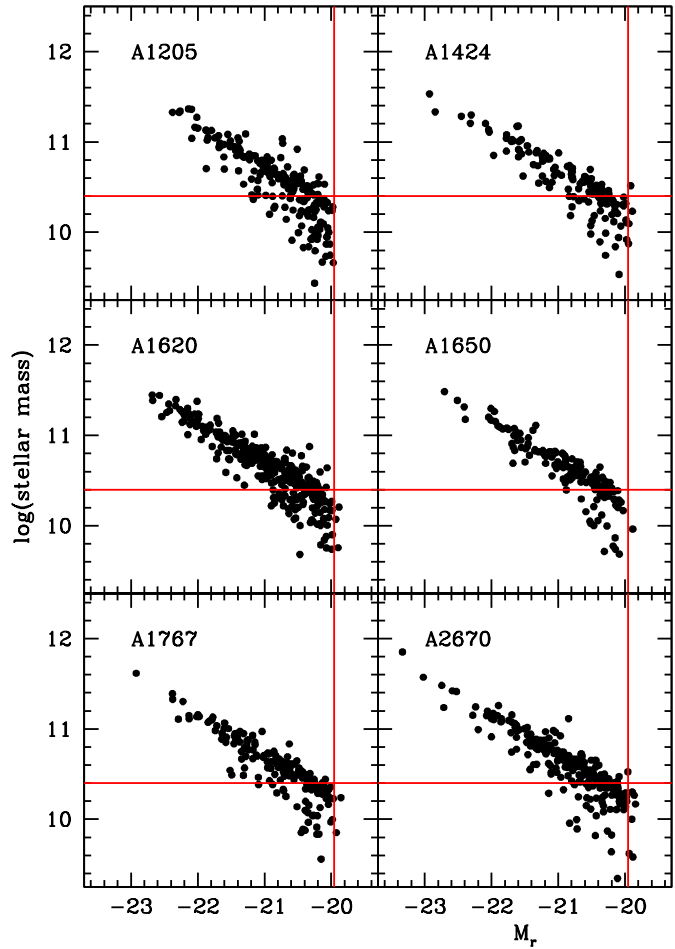,angle=0,width=5.5in}} 
\vspace*{-0.5cm}
\caption{Absolute r-band magnitude versus galaxy stellar mass (expressed as
the logarithm of the solar mass of the galaxy) for the clusters in 
our sample. The vertical line denotes the spectroscopic limit of $r=17.77$ in
our most distant cluster (Abell~1620) and the 
horizontal line denotes $\log({\mathrm stellar~mass})=10.4$
that corresponds to the mass limit that we are $\sim$complete to.
These lines are replicated for the other 5 clusters.  We only use
galaxies that are more massive and brighter than these limits for our subsequent analysis
to avoid biasing our sample.
}
        \label{fig:bias}
\end{figure}

To create our final sample, we stack all of our 
clusters together to form a composite sample. Analogous to Pimbblet (2011), this
is achieved by placing the clusters on to a common scale (i.e.\ $R_{\mathrm virial}$) and
limiting our clusters to a common absolute magnitude (i.e.\ the absolute magnitude
corresponding to $r=17.77$ -- $M_r=-19.96$ -- at the redshift of our most distant cluster, Abell~1620; 
this corresponds to approximately $M^{\star}+1$ along the 
luminosity function according to the analysis of Jensen \& Pimbblet 2012). 
As the terminal step, we select a mass limit for our clusters to prevent
our sample being biased from having a low $r$-band limit that is coupled
with a top-end mass limit (Holden et al.\ 2007; see also
Pimbblet \& Jensen 2012).  To achieve this, we examine plots of absolute magnitude
versus stellar mass for each of our clusters and restrict our sample to those
galaxies brighter than $M_r=-19.96$ and more massive than 
the most massive galaxy $\log({\mathrm stellar~mass})=10.4$ at this limiting magnitude for our most distant 
cluster (Fig.~\ref{fig:bias}).

The final, bias-corrected composite sample consists of 300 galaxies within $R_{\mathrm virial}$, and
701 galaxies within $3 R_{\mathrm virial}$ from these 6 clusters.

\subsection{Sample Validation}
Before proceeding with our analysis, we elect to perform some 
validation tests on our sample to ensure that they are free
of substructure -- at least in their centres -- as we have suggested.
The need for such a validation step is our concern that the literature
expresses both ambiguous and conflicting statements about some of
the clusters in our sample.  For example, Abell~1650 is 
noted by Einasto et al.\ (2012) as being a unimodal cluster, in 
agreement with Pimbblet et al.\ (2002) who term it a morphologically
regular cluster using a large radii dataset.  
This is in contrast to Flin \& Krywult (2006)
who note Abell~1650 has `substructures in the cluster field'.

Taking our bias-corrected sample, we apply the Dressler \& Schectman
(1988; DS) test for substructure to each cluster individually.
The DS test is probably the most powerful available to detect substructure 
in arbitrary three dimensional datasets (Pinkney et al.\ 1996) and 
we therefore consider it perfectly adequate for our sample validation check.
Briefly, the DS test works by finding the local mean velocity and
standard deviation of the 10 nearest neighbours to a given galaxy 
and compares them to the global values for the cluster, such that:
\begin{equation}
\delta^2 = \left( \frac{ N_{\mathrm local} + 1 }{\sigma_{\mathrm global}^2} \right)
[ (\overline{cz}_{\mathrm local}-\overline{cz}_{\mathrm global})^2 + ( \sigma_{\mathrm local}-\sigma_{\mathrm global})^2 ]
\end{equation}

where the parameter $\delta$ yields a measure of the deviancy of
this sub-sample.  A summed parameter of merit, $\Delta$ is then 
computed by summing all $\delta_i$ terms in each cluster. To get
a handle on the probability of $\Delta$ occurring, the velocity data are
shuffled randomly between member galaxies 1000 times in a Monte Carlo fashion
and the actual value of $\Delta$ is then compared to this ensemble.

In Table~\ref{tab:ds}, we give the values of $P(\Delta)$ for the clusters
in our sample.  All of our clusters are substructure free within $1 R_{\mathrm virial}$
to the limits probed by our investigation.
However, Abell~1620 stands out from the others 
as possessing significant substructure at high radii from the 
cluster centre.  This agrees with the analysis of Burgett et al.\ (2004)
who also suggested that the cluster may contain substructure from
a complementary analysis of 2dF data. We analyze this cluster further 
in Appendix~A.
Despite the substructure at high cluster-centric radii, we retain Abell~1620
in our sample.  We have experimented with removing this cluster
from our sample and find that the effect on
our primary results are negligible, but our uncertainties become fractionally 
larger. 

Finally, we issue the caveat that even though this simplistic test
has indicated no substructure in the centres of our clusters, this does
not preclude substructure at fainter magnitude limits arising from
coherent, potentially low mass, infalling groups as might be expected 
(cf.\ Owers et al.\ 2011).

%
%
\begin{table}
\begin{center}
\caption{Results of the DS test using various cuts in radii.
Results under 0.01 are considered to indicate significant substructure.
\hfil}
\begin{tabular}{llllll}
\hline
Cluster & \multicolumn{3}{c}{$P(\Delta)$} \\
        & $r<R_{\mathrm virial}$  &  $r<2R_{\mathrm virial}$  &  $r<3R_{\mathrm virial}$ \\
\hline
A1205   &  0.39 & 0.12 & 0.08 \\
A1424   &  0.20 & 0.03 & 0.05 \\
A1620   &  0.03 & $<0.01$ & $<0.01$ \\
A1650   &  0.05 & 0.05 & 0.21 \\
A1767   &  0.38 & 0.34 & 0.33 \\
A2670   &  0.25 & 0.31 & 0.29 \\
\hline
\vspace*{-1.0cm}
\end{tabular}
  \label{tab:ds}
\end{center}
\end{table}

\subsection{AGN identification}
\begin{figure}
\centerline{\psfig{file=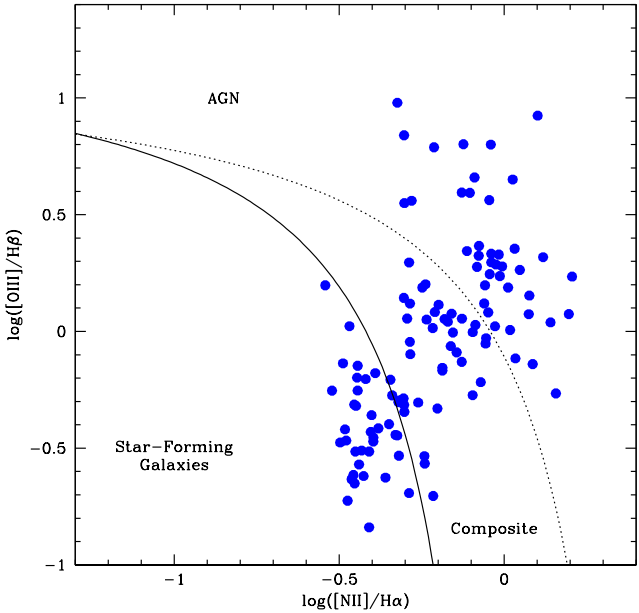,angle=0,width=3.5in}}
\caption{BPT plane for our composite cluster galaxy sample.  
All points have S/N$>3$ in each line.
The solid curve is the Kauffmann et al.\ (2003) demarcation line:
galaxies above the curve are designated AGN, those below are
simply regular star-forming galaxies. Also shown is the Kewley et al.\ (2001)
demarkation curve (dotted line). Galaxies between the two curves are composites:
weaker AGN whose hosts are also star-forming.
}
        \label{fig:bpt}
\end{figure}

In order to identify which galaxies in our sample are AGN, we make
use of a BPT diagram (Baldwin, Phillips \& Terlevich 1981; 
see also Veilleux \& Osterbrock 1987).
The BPT plane consists of flux ratio of [NII]$_{\lambda6583} / H\alpha$
versus [OIII]$_{\lambda5007} / H\beta$.  
The measurements of equivalent widths for these
lines are taken from Tremonti et al.\ (2004; 
see also www.mpa-garching.mpg.de/SDSS).
In Fig.~\ref{fig:bpt} we plot the position of all galaxies
in the composite cluster on the BPT plane that have S/N$>3$ in the 
necessary lines. To differentiate AGN from galaxies that are simply
star-forming, we use the demarcation curve of Kauffmann et al.\ (2003).
The curve is a refinement of earlier work by Kewley et al.\ (2001)
and yields 30 AGN within $R_{\mathrm virial}$ and 81 AGN within $3 R_{\mathrm virial}$.
We also define a composite sample -- those galaxies that lie between
the Kaufmann and Kewley curves -- these galaxies are weaker AGN whose
host galaxies are star-forming. By implication, the use of the BPT
diagram to select AGN means that our sample are composed of 
`cold-mode' AGN (cf.\ Kere{\v s} et al.\ 2005; Hopkins
\& Hernquist 2006).

From the outset, we note that there is no significant optical colour
difference (e.g.\ in $g-r$) between our AGN sample and the rest
of the cluster population. This holds true even if we divide our sample
by galaxy stellar mass.

\section{AGN fraction}
We now compute the cluster AGN fraction in two ways: by radius from
the centre of the cluster and by galaxy mass.  
We note that the stellar masses of the galaxies
are from the SDSS value-added catalogue (www.mpa-garching.mpg.de/SDSS; 
see also Kauffmann et al.\ 2003).

\subsection{Fraction by Radius from Cluster Centre}
\begin{figure}
\centerline{\psfig{file=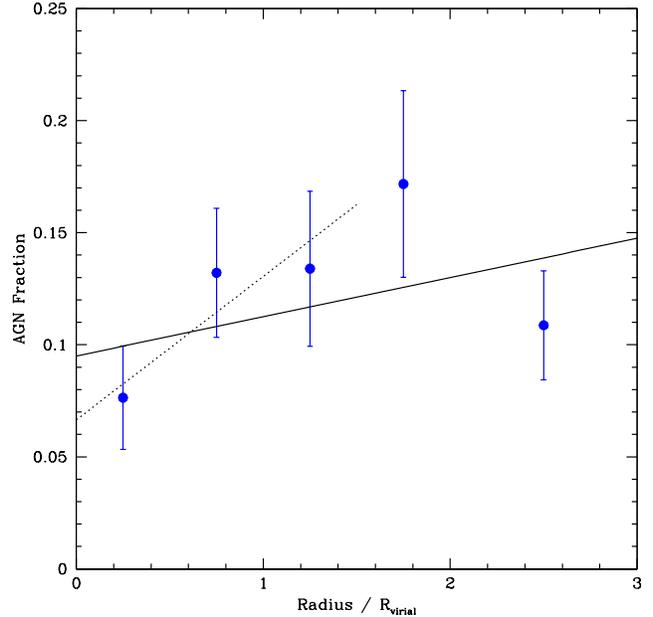,angle=0,width=3.5in}}
\caption{AGN fraction in the composite cluster as a function
of radius from the centre, with Poisson errors.  Each bin covers
$0.5 R_{\mathrm virial}$ except the last bin that covers $1 R_{\mathrm virial}$
to ensure that each point in the plot has $>100$ member galaxies (cf.\ 
Cameron 2011).
A line of best fit to these points is overplotted as the solid line.
This line has a gradient of $d({\mathrm fraction})/d R_{\mathrm virial} = 0.018 \pm 0.020$.
We also display a line of best fit to the three points within $1.5 R_{\mathrm virial}$
(dotted line), which has a gradient of 
$d({\mathrm fraction})/d R_{\mathrm virial} = 0.064 \pm 0.021$.
}
        \label{fig:agn1}
\end{figure}

From the outset of our analysis, we note that we have selected
the brightest cluster member as being the centre of our galaxy
clusters.  Although other choices could have been made, 
such as a luminosity-weighted centre or the
peak X-ray flux location, we note that varying this choice does
not alter the primary results presented in this work.
We compute the AGN radial fraction in terms of $R_{\mathrm virial}$ (a
more physically meaningful scale than 
a fixed metric that uses Mpc; cf.\ Pimbblet et al.\ 2002) 
and plot the result in Fig.~\ref{fig:agn1}.  
The AGN fraction is found to increase with distance from
the cluster centre at a rate of $d({\mathrm fraction})/d R_{\mathrm virial} = 0.018 \pm 0.020$.

This rate of increase is not significant, which 
may explain some of the differences reported in the 
literature concerning this fraction (for instance, 
contrast Gilmour et al.\ 2007
and Kauffmann et al.\ 2004 with Miller et al.\ 2003
and Martini et al.\ 2007).
That said, if we restricted our analysis to the three 
innermost points of Fig.~\ref{fig:agn1} (i.e.\ $<1.5 R_{\mathrm virial}$)
we find a significant gradient of 
$d({\mathrm fraction})/d R_{\mathrm virial} = 0.064 \pm 0.021$.
For those studies that split the radial AGN fraction into
bins similar to our analysis (Coldwell et al.\ 2002), similar
results are obtained.  

Further, if we were to compare the AGN fraction within 
$1.5 R_{\mathrm virial}$ to that at 2.0 -- 3.0 $R_{\mathrm virial}$,
we find no significant difference (AGN fractions
of $0.11 \pm 0.02$ and $0.11 \pm 0.03$, respectively).
This may explain why investigations that 
that compare the `cluster' environment 
to a control `field' may be biased to finding
no difference in AGN fraction. 
Moreover, the types of cluster used can also bias the
measurement of AGN fraction --
those investigations that use all types
of cluster to probe AGN fraction 
could be biased by the presence of substructure 
-- a bias that the present work intentionally avoids -- 
as could the use of very high central density clusters versus low density.

The lines of best fit presented in Fig.~\ref{fig:agn1} could
also be over-simplifications of the situation.  For example,
Porter et al.\ (2008) report that there is an enhancement
of specific star formation rate at $\sim$few Mpc away from
cluster centres caused by harassment inside galaxies being
accreted along filaments of galaxies (see also
Koyama et al.\ 2008; Perez et al.\ 2009). 
If true, then we may
expect a similar enhancement of AGN fraction just beyond the
virial radii of our clusters.  Although Fig.~\ref{fig:agn1}
displays a local maxima in AGN fraction at $1.75 R_{\mathrm virial}$,
it is not significant -- a larger sample of clusters 
and bona-fide filaments will be required to fully address 
this question.

\subsection{Fraction by Mass}

\begin{figure}
\centerline{\psfig{file=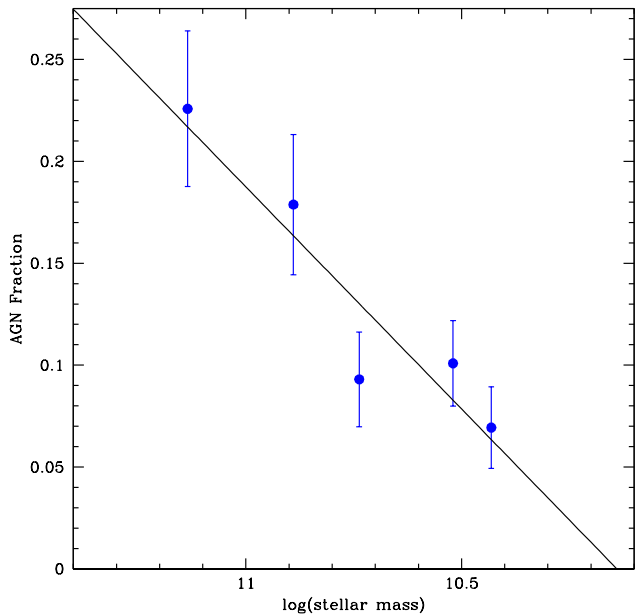,angle=0,width=3.5in}}
\caption{As for Fig.~\ref{fig:agn1}, but as a function
of galaxy stellar mass; each point has the same number of galaxies.  The line
of best fit has a gradient of 
$d({\mathrm fraction})/d \log({\mathrm stellar~mass}) = 0.22 \pm 0.05$ across the range studied.
}
        \label{fig:agn2}
\end{figure}

In Fig.~\ref{fig:agn2} we 
compute the AGN fraction as a function of 
galaxy stellar mass for all galaxies within
$3 R_{\mathrm virial}$.  We fit this data with a line of
best fit and find that it has a significant gradient 
of $d({\mathrm fraction})/d \log({\mathrm stellar~mass}) = 0.22 \pm 0.05$ --
higher mass galaxies are significantly more likely to
host AGN than their lower mass cousins (cf.\ Tanaka 2012;
Pimbblet \& Jensen 2012; Xue et al.\ 2010; Brusa et al.\ 2009; 
Floyd et al.\ 2004; Dunlop et al.\ 2003;
see also Best et al.\ 2005).

\begin{figure}
\centerline{\psfig{file=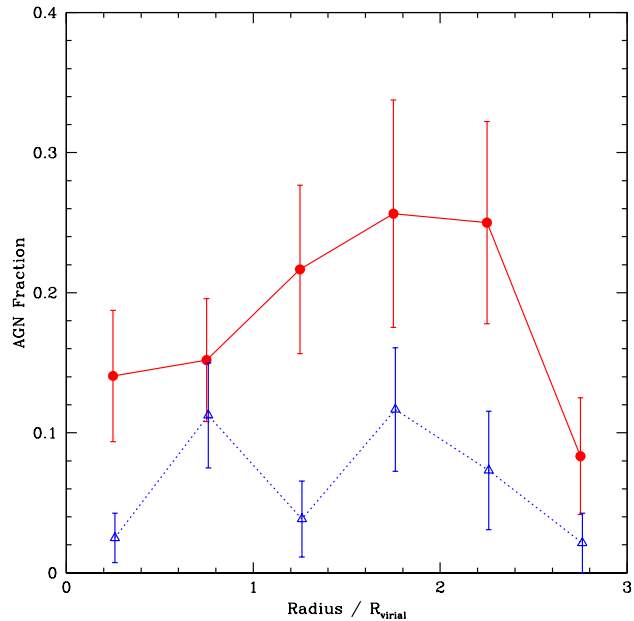,angle=0,width=3.5in}}
\caption{As for Fig.~\ref{fig:agn1}, but for galaxies with
$\log({\mathrm stellar~mass})>10.7$ (filled red circles, solid line) and
$<10.7$ (open blue triangles, dotted line). The higher mass regime contains
a higher fraction of AGN at all radii.
}
        \label{fig:agn3}
\end{figure}

To examine if a mass selection would affect the radial
cluster AGN fraction, we repeat our above radial analysis 
for two mass bins in Fig.~\ref{fig:agn3}, split 
(arbitrarily) at $\log({\mathrm stellar~mass})=10.7$ to ensure
that there are approximately equal numbers of galaxies
above and below that mass in our sample.
The more massive galaxies have a larger AGN fraction
at all radii and there is a steady
(but not very significant)
increase in AGN from the cluster center to $2R_{\mathrm virial}$
before it drops slightly lower again. Conversely, the lower
mass galaxies do not vary in fraction significantly.
The deficit of AGN in the centre of clusters may therefore 
simply be a reflection of the changing mix of galaxy types 
(e.g.\ colour; morphology; mass) with
cluster radius (cf.\ von der Linden et al.\ 2010).
This is illustrated in Table~\ref{tab:highmass}, where
we note the fraction
of galaxies with $\log({\mathrm stellar~mass})>11.0$ 
in our sample steadily decreases with radius from the centre
of our stacked cluster.
This covariance of radius with mass is simply an expression of
the well-known morphology-density relation (e.g., Dressler 1980;
Dressler et al.\ 1997; Smith et al.\ 2005; see also Baldry et al.\ 2006).

%
%
\begin{table}
\begin{center}
\caption{Illustration of how the
fraction of our most massive galaxies ($\log({\mathrm stellar~mass})>11.0$) in our sample 
changes as a function of radius.
\hfil}
\begin{tabular}{ll}
\hline
Radius & Fraction \\
($R_{\mathrm virial}$) & \\
\hline
0--1 & $0.19 \pm 0.02$ \\
1--2 & $0.16 \pm 0.03$ \\
2--3 & $0.14 \pm 0.03$ \\ 
\hline
\vspace*{-1.0cm}
\end{tabular}
  \label{tab:highmass}
\end{center}
\end{table}

\section{Discussion}

\subsection{AGN Phase-Space and Local Galaxy Density}

\begin{figure}
\centerline{\psfig{file=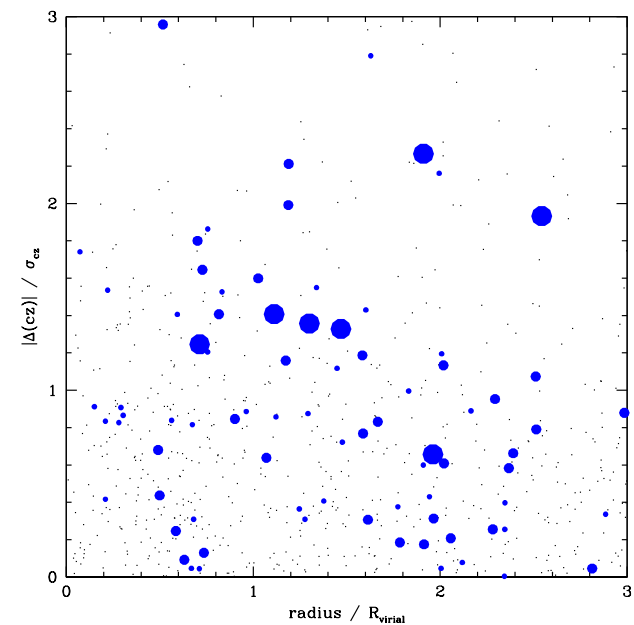,angle=0,width=3.5in}}
\caption{Phase-space diagram of AGN
and other cluster members (dots) on a $\Delta(cz)$ versus
radius plane.  The AGN are coded according to their [OIII] emission.
The largest blue points are AGN with an equivalent width of [OIII] $<-10$ \AA;
the medium sized points have [OIII] equivalent widths in the range $-1.5$ -- $-10$~\AA; 
all the other (smallest) blue points have [OIII] equivalent width $>-1.5$~\AA.
}
        \label{fig:phase}
\end{figure}

%
%
\begin{table*}
\begin{center}
\caption{Fraction of retired AGN and galaxies from the total
population taken from Fig.~\ref{fig:cid} as a function of
radius and galaxy mass.  The mass bins are chosen to have
approximately equal galaxy numbers in the total population.
\hfil}
\begin{tabular}{lll}
\hline
Bin             & Retired       & Retired fraction \\
                & AGN fraction  & of total population \\
\hline
0--1 $R_{\mathrm virial}$  & $0.45\pm0.12$ & $0.37\pm0.04$ \\
1--2 $R_{\mathrm virial}$  & $0.34\pm0.10$ & $0.36\pm0.04$ \\
2--3 $R_{\mathrm virial}$  & $0.20\pm0.10$ & $0.41\pm0.05$ \\
\hline
$\log({\mathrm stellar~mass})>10.809$     & $0.51\pm0.10$ & $0.53\pm0.05$ \\
$10.575<\log({\mathrm stellar~mass})<10.809$ & $0.12\pm0.08$ & $0.34\pm0.04$ \\
$\log({\mathrm stellar~mass})<10.575$     & $0.12\pm0.08$ & $0.26\pm0.03$ \\
\hline
\end{tabular}
  \label{tab:retired}
\end{center}
\end{table*}

If AGN fractions are being enhanced in the cluster
outskirts by interactions with other galaxies,
then an investigation of their locations in $(cz-\overline{cz})/\sigma_{cz}$
versus radius / $R_{\mathrm virial}$ phase-space may reveal this.
This phase-space is plotted in Fig.~\ref{fig:phase}.
We test if the AGN and other galaxy populations are distributed differently 
on this plane through a two-dimensional K-S test (Fasano 
\& Franceschini 1987; Peacock 1983) and find that the two populations
are the same -- i.e.\ the AGN are no more likely to be at large values
of $|(cz-\overline{cz})|/\sigma_{cz}$ plus radius/$R_{\mathrm virial}$
(or conversely, centrally concentrated)
than other cluster galaxies.  This holds even if we consider 
only galaxies more massive (or less massive) than
$\log({\mathrm stellar~mass})=10.7$. Indeed, both the mean
and median values of $|(cz-\overline{cz})|/\sigma_{cz}$
for the AGN and general cluster population 
are within $1\sigma$ of each other; even within individual 
$1R_{\mathrm virial}$ radial bins.  

\begin{figure*}
\centerline{\psfig{file=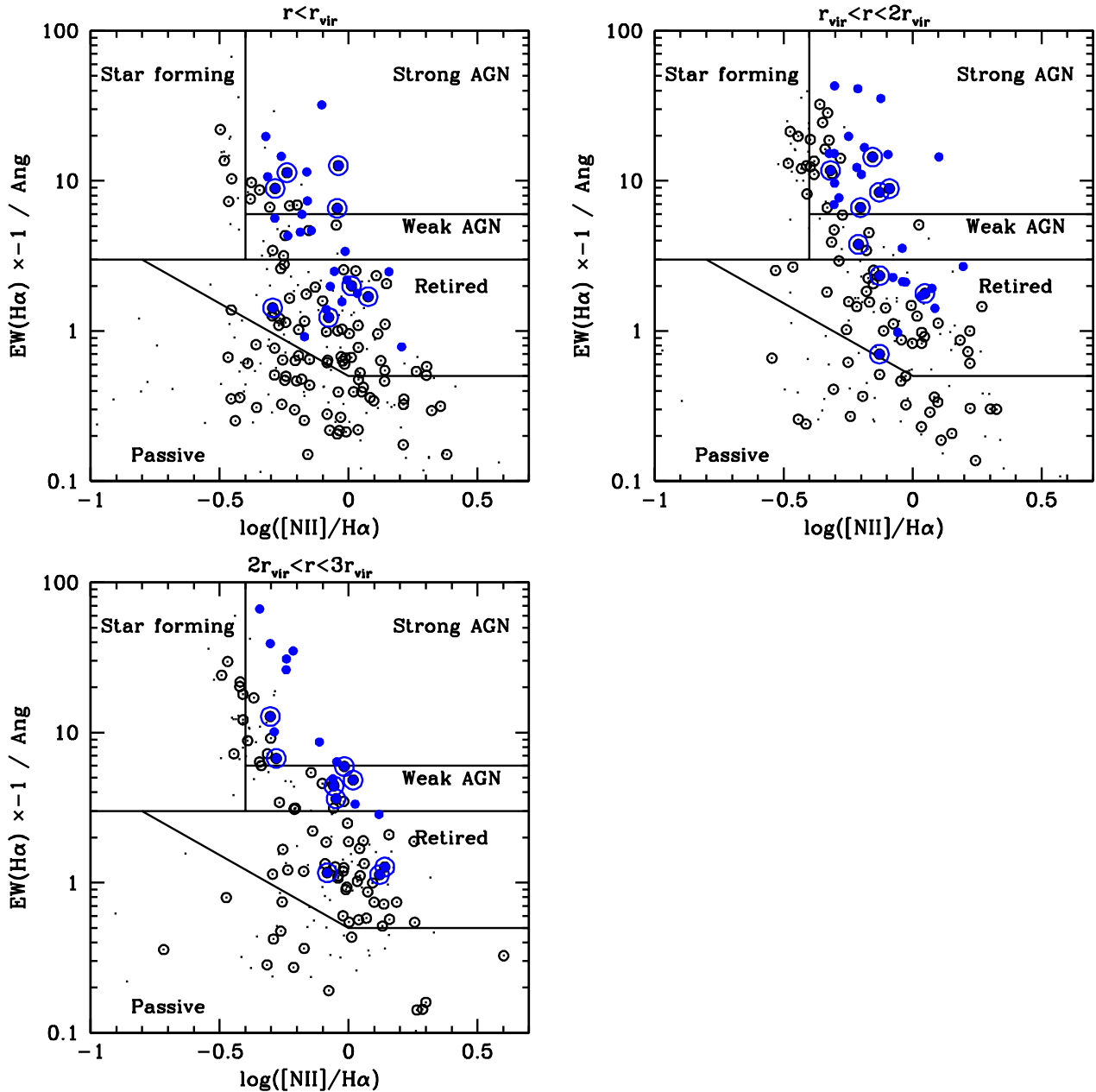,angle=0,width=7.25in}}
\vspace*{-1cm}
\caption{Diagnostic plot of Cid Fernandes et al.\ (2011) applied to
our sample and split by radius.
The larger blue points are our AGN, as defined by the Kauffmann
demarkation line.  The small dots are all our other galaxies, regardless of
whether they have a high S/N ratio (i.e.\ unlike the definition of our
AGN sample).  Circled points denote low velocity offset galaxies with
$|\Delta (cz)| / \sigma_{cz} < 0.5$.  Although $\sim$35 per cent of our
AGN may be retired galaxies under this classification, those AGN
with low radii and velocity offset are not preferentially retired.
}
        \label{fig:cid}
\end{figure*}

These results are in apparent disagreement with Haines et al.\ (2012;
see especially their Fig.~6)
who find that X-ray selected AGN are significantly more likely to 
be at the cluster infall regions than the general cluster population.
Apart from using X-ray selected AGN, Haines et al.\ (2012) also use
different cluster selection criteria: their clusters are more massive
than the ones employed here and at higher redshift -- the mean difference
in lookback time between our sample and Haines et al.~is $\approx1.7$~Gyr --
and the clusters may possess significant subclustering even at bright 
magnitudes.  
On the other hand, finding several AGN at comparatively low velocity offsets and
radii is not absolute proof against them being an infalling population, as they can
easily still appear at these locations (see Fig.~10 of Haines et al.\ 2012). 
An alternative hypothesis is that the AGN at low radii and velocity offsets
are `retired galaxies' -- i.e.\ galaxies 
whose ionization mechanism is provided by old stellar populations
(e.g.\ Cid Fernandes et al.\ 2010; 2011; Yan \& Blanton 2012 and references therein).
In Fig.~\ref{fig:cid} we plot our sample using the WHAN diagnostic plot of
Cid Fernandes et al.\ (2011). This plot is more `efficient' than the standard BPT
approach as it only uses two lines: H$\alpha$ and [NII]. Moreover, it is readily
able to disentangle the so-called `retired' galaxy population from weak AGN 
types. The WHAN diagram can also classify up to 50 per cent more of the emission
line galaxy types than the BPT approach, and is therefore more able to distinguish
LINERs from Seyferts, but no information from the BPT is
`lost' in the move to the WHAN diagnostic.

There are a number of noteworthy aspects of Fig.~\ref{fig:cid}. At a basic
level, it reflects already well-known results that actively star-forming galaxies
reside at the outskirts of clusters (i.e.\ there are fractionally fewer circled dots in
the star-forming corner 
compared the passive population), whereas the passive galaxies dominate the low velocity offset,
low radii population (cf.\ Pimbblet et al.\ 2006; Pimbblet \& Jensen 2012 and references
therein).  Of our AGN population, some $35\pm6$ per cent fall in to the `retired' classification.
These AGN are not preferentially situated at low radii and low velocity offsets:
only $36\pm10$ per cent of the retired AGN satisfy such a criteria.  
This is, however, a slightly
larger fraction than the other AGN classes: 
$14\pm10$ per cent of weak AGN and $16\pm6$ per cent of strong AGN reside at low radii 
plus low velocity offsets. These statistics are not significant enough to infer 
a duty cycle, but it is clear that all classes of AGN reside at all positions in 
our clusters.

We interrogate Fig.~\ref{fig:cid} to determine how the retired fraction 
of both AGN and the entire galaxy population varies as a function of radius from
the cluster centre and galaxy mass and present these results in 
Table~\ref{tab:retired}. Whilst the retired fraction of all galaxies is approximately
constant with radius, the retired fraction of AGN mildly decreases with distance
away from the centre of our clusters.  The trend with mass is steeper: high mass galaxies
are much more likely to be in the retired class than lower masses.
This result agrees with the contention of Lee et al.\ (2012) that the presence of
a bar in a galaxy is no trigger for AGN activity (see also Combes 2003).  Rather, the
presence of bars and AGN is simply driven by host galaxy mass 
(e.g., Sheth et al.\ 2008; Nair \& Abraham 2010).

Taking Fig.~\ref{fig:phase} and Fig.~\ref{fig:cid} together, there is some suggestion
that the disagreement with Haines et al.\ (2012) outlined above may not be so serious.
By considering only the those AGN with equivalent widths of [OIII]$<-10$~\AA, we see
that they seem to lie above $|\Delta(cz)|/\sigma_{cz}=1.2$ (Fig.~\ref{fig:phase}).
This is reflected in Fig.~\ref{fig:cid}:
those AGN with the highest H$\alpha$ emission values have larger 
velocity offsets (i.e.\ those points that are not circled in Fig.~\ref{fig:cid}). 
This holds in all three panels of Fig.~\ref{fig:cid} -- i.e.\ at all radii.
Haines et al.\ (2012) notes that their (powerful, X-ray) AGN 
reside on infalling caustic; 
hence if we only examine those powerful AGN in our sample, 
we come to an analogous conclusion.

We can also examine if the AGN are preferentially in
areas of high galaxy density by computing the local galaxy
density for each galaxy in our sample.  We choose $\Sigma_{5}$
-- the surface area on the sky that is occupied by a given
galaxy to its tenth nearest neighbour --
as our estimator of local galaxy density; this
is effectively a probe of the internal densities of the dark matter halos
(Muldrew et al.\ 2012). 
The plot of $\Sigma_5$ as a function of 
radius from the cluster centre is shown in Fig.~\ref{fig:sigma}.
There are a number of obvious AGN at high local galaxy density, but
to test whether the AGN are at a systematically higher value of
$\Sigma_5$ we compare bootstrapped
mean $\Sigma_5$ values of the AGN against
the rest of the cluster members as a function of radius (Table~\ref{tab:med}).
This shows no significant difference between the two populations at
any radii.

Therefore, if AGN are being triggered by encounters with other galaxies,
they have since moved away from the site of the interaction suggesting
a suitably long timescale between interaction and subsequent AGN enhancement.
Detailed computation of the value of this time-lag is beyond the scope of
the present work, but we would expect it to be less than 
the time required for substructure to homogenize (e.g., Araya-Melo et 
al.\ 2009).  
Other studies (e.g.\ Schawinski et al. 2007; Shabala et al.\ 2012)
have found that when interactions trigger AGN, it takes $\sim$100 -- 200 Myr 
for an AGN to ``switch on''. 
Given a velocity dispersion of $\sim$1000 km/s (cf.\ Table~\ref{tab:clusters}), 
each galaxy could move 100 -- 200~kpc in this time. 
Therefore we regard it as comparatively easy to 
``wipe out'' the local density enhancement signature by the time the AGN switches on.

To round off this part of our discussion, we now cross-match our sample with 
the FIRST radio database to ascertain which of our galaxies would be classified
as radio AGN (or `hot-mode') and determine their radial fractions. 
We find 28 matches from FIRST using a cross-matching radius of 10 arcsec to our sample.
To see whether the detected radio emission can arise due to just star formation,
we examine the SFR values for these galaxies from the value-added SDSS catalogues.  
We convert the reported SFR
to a 1.4 GHz radio luminosity using the formula of Yun et al.\ (2001) and determine
how 
much higher the FIRST radio luminosity was than the value expected purely from star formation.
Radio AGN are then extracted as those galaxies whose FIRST radio luminosity exceeds
that expected from the star formation within the SDSS fibre by a factor $>1\sigma$ 
(where $\sigma$ is the standard deviation in the SFR estimate).  This results in 15 radio 
AGN. This sample has a markedly high mass: the median is 
$\log({\mathrm stellar~mass} / M_\odot)=11.2 \pm 0.3$.
The radio AGN fraction for these galaxies goes as $0.013\pm0.007$, 
$0.028\pm0.011$, $0.027\pm0.012$ for bins of $1R_{\mathrm virial}$ from the cluster centre.
These small number statistics are hard to draw a meaningful conclusion from

%
%
\begin{table}
\begin{center}
\caption{Bootstrapped means and standard deviations 
of $\Sigma_5$ values for the AGN and other
cluster members as a function of radius from the cluster 
centre. The two samples are statistically drawn from 
the same parent population.
\hfil}
\begin{tabular}{lll}
\hline
Radius		& AGN $\Sigma_5$ & Other $\Sigma_5$ \\
($R_{\mathrm virial}$)  & (Mpc$^{-2}$)    & (Mpc$^{-2}$)      \\
\hline
0--0.5          & $56.4\pm13.9$ & $39.1\pm5.6$   \\
0.5--1.0        & $18.4\pm2.7$  & $19.2\pm1.5$  \\
1.0--1.5        & $35.2\pm20.8$  & $26.1\pm5.3$  \\
1.5--2.0        & $4.3\pm0.9$  & $5.4\pm0.6$   \\
2.0--2.5        & $8.8\pm3.2$  & $10.8\pm2.2$  \\
2.5--3.0        & $3.0\pm1.5$  & $4.1\pm1.7$  \\
\hline
\end{tabular}
  \label{tab:med}
\end{center}
\end{table}

\subsection{Composite AGN and AGN Power}
In Fig.~\ref{fig:bpt}, we identified not only the AGN (those galaxies
above the Kaufmann demarkation line), but also a sample of composite
weak AGN and star-forming galaxies (those between the Kaufmann and Kewley
demarkation lines).
In Fig.~\ref{fig:kewkauf} we divide the radial AGN fraction in to the
Kewley demarcated AGN, and the composite sample to ascertain
if the composite sample are driving
any of the trends seen above. The composite sample appears to
follow quite a flat distribution.  Meanwhile, the Kewley et al.\ (2001) 
defined AGN show a steeper initial variation with increasing radius
which plateaus quickly.  Both samples are consistent with the
trend depicted in Fig.~\ref{fig:agn1} for the AGN fraction gradient.
However, if we consider only the inner points (i.e.\ the dotted line
in Fig.~\ref{fig:agn1}), we see that this is more consistent with the
Kewley defined AGN.

\begin{figure}
\centerline{\psfig{file=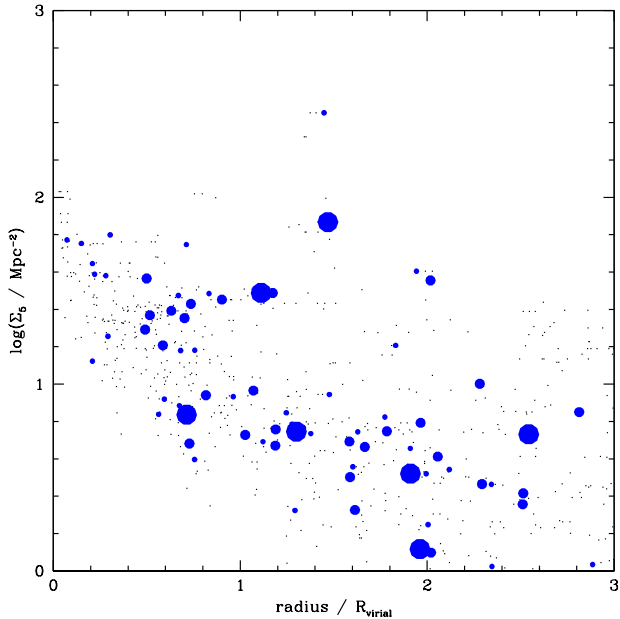,angle=0,width=3.5in}}
\caption{Local galaxy density, $\Sigma_5$ as a function of
radius from the cluster centre.  The large blue points denote
the AGN, coded as per Fig.~\ref{fig:phase};
small dots are the other galaxies. The AGN statistically
occupy the same region of this parameter space as the other cluster
members (Table~\ref{tab:med}).
}
        \label{fig:sigma}
\end{figure}

\begin{figure}
\centerline{\psfig{file=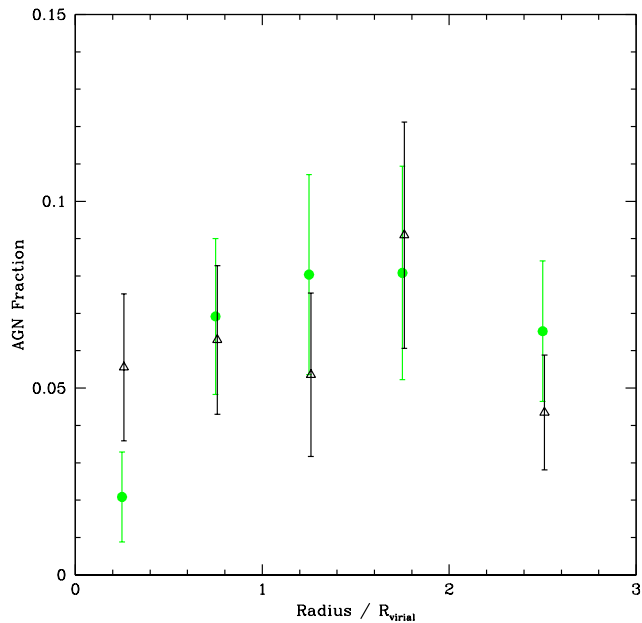,angle=0,width=3.5in}}
\caption{As for Fig.~\ref{fig:agn1}, but split on the basis
of whether the galaxies falls in the `composite' AGN+star-forming
sample (black open triangles), or the Kewley et al.\ (2001) definition
of AGN (green filled circles). The two samples are fractionally offset in
radius from each other for clarity.  
}
        \label{fig:kewkauf}
\end{figure}

If there is a bona-fide radial trend of AGN fraction with radius,
we may be able to see this reflected in the [OIII] line 
strength which is a proxy for AGN power.  In Table~\ref{tab:o3}, we
list the median equivalent widths of [OIII] 
for our original
AGN (i.e.\ Kauffmann delineated) with radius (see also
the different point sizes in Fig.~\ref{fig:phase} and Fig.~\ref{fig:sigma}).  
For comparison,
we also detail the same measurement for all galaxies in our sample.
No significant radial trend is observed for either sample.  
In the case of the AGN, this may simply be because
we lack the numbers to detect such a trend since we only have 84
AGN within $3 R_{\mathrm virial}$ (and a maximum of 32 galaxies in a 1Mpc
bin in Table~\ref{tab:o3}).  A larger sample will be required
to investigate this.
We can, however, infer that we are not missing any 
AGN because they are simply too weak 
and therefore below the BPT detection threshold.
Hence the change in AGN fraction is a bona-fide change in the duty cycle 
rather than AGN luminosity.

%
%
\begin{table}
\begin{center}
\caption{Variation of [OIII] equivalent width (EW) with radius.
The quoted uncertainties are bootstrapped standard deviations.
\hfil}
\begin{tabular}{lll}
\hline
Radius & AGN Sample       & All Galaxies \\
($R_{\mathrm virial}$)  & Median [OIII] EW & Median [OIII] EW \\
       & (\AA)            & (\AA) \\
\hline
0--1 & $-1.3 \pm 0.2$ & $-0.4 \pm 0.1$ \\
1--2 & $-1.6 \pm 0.7$ & $-0.5 \pm 0.1$ \\
2--3 & $-1.6 \pm 0.7$ & $-0.6 \pm 0.1$ \\
\hline
\end{tabular}
  \label{tab:o3}
\end{center}
\end{table}

\subsection{AGN Colour}
\begin{figure*}
\centerline{\psfig{file=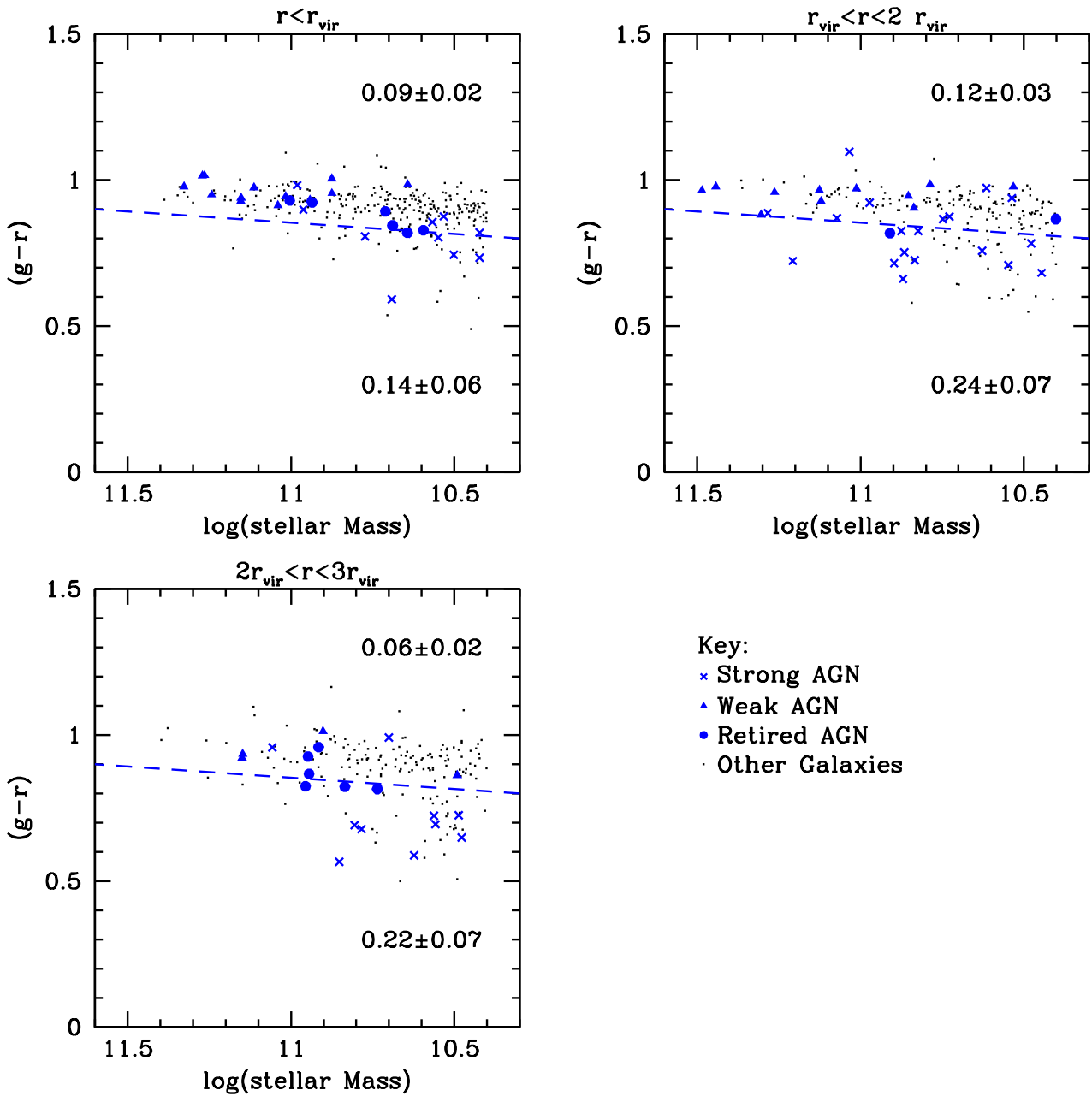,angle=0,width=7.25in}}
\vspace*{-1cm}
\caption{Colour-mass diagrams for our sample, split by radius 
to the cluster centre. AGN in our sample have been marked according
to their WHAN classification from Fig.~\ref{fig:cid} (crosses for strong AGN; 
filled triangles for weak AGN; filled circles for the retired class).
The dashed diagonal line is our approximation for the division between
red sequence (above the line) and blue cloud galaxies (below the line).
The numbers show the corresponding AGN fractions above and below this line.
These fractions remain statistically constant with radius.
}
        \label{fig:colmass}
\end{figure*}

If AGNs are preferentially
associated with bright galaxies that are presumably morphologically late-type, 
then determining the fraction of
AGNs in radial bins in our sample should mimic the radial
distribution of late-type galaxies in clusters -- i.e.\ the morphology-density 
relation (Dressler 1980; see above) and therefore not address whether
a specific mechanism is tied to AGN activation or quenching.
To partially resolve this, we could attempt to morphologically classify
all of our sample through using (e.g.) GalaxyZoo (Lintott et al. 2008) in an analogous
way to Pimbblet \& Jensen (2012).
But this approaches contain problems: principal among them being
a large fraction of `uncertain' classifications that could skew an analysis.
To combat this, we construct a colour-mass diagram for our sample, divided 
by radius to the cluster centre (Fig.~\ref{fig:colmass}).
We divide this diagram up in to two halves: a red sequence and a blue cloud
component.  This is done by eye, choosing a line that divides the two reasonably
cleanly.  Despite the arbitrary nature of this approach, it serves our purpose
of creating two categories of galaxies (i.e.\ early and late types) and we
note that the gradient is consistent with
the colour-magnitude relations presented in earlier works (Pimbblet et al.\ 2006; 2002)
and the lower envelope limit of such fitted colour-magnitude relations. 
However, we caution that we have made no attempt to correct the (g-r) colours in
Fig.~\ref{fig:colmass} for AGN blueing: given the use of SDSS model magnitudes,
we suggest that this effect would be small and only likely to affect the strong
(i.e.\ bluest) AGN population since the weak and retired AGN are already 
predominantly residing on the red sequence (Fig.~\ref{fig:colmass}).   

From Fig.~\ref{fig:colmass}, we see that 
any enhancement of the red sequence AGN fraction at low radii appears to be purely 
driven by massive galaxies.  We quantify this in Table~\ref{tab:colmassfracs}
where we detail the AGN fractions not only above and below the red sequence 
envelope cut-off, but also divided by mass. Although the uncertainties on these
numbers are large (too large to infer statistically significant trends), 
by considering galaxies lying within and outside 1~$R_{\rm virial}$ of the cluster centre,
it is tempting to speculate that red massive AGN and 
blue low-mass AGN may have a radial dependence whereas the red sequence low mass
AGN may not. If so, this may imply a common AGN triggering mechanism such as
gas-rich interaction.  
Hence if a low-mass galaxy underwent such an interaction, 
it would necessarily become blue due to the parallel star formation. 
This would not be the case for a massive galaxy 
since it will have a lower specific star formation rate which in turn would 
correlate with the (g-r) colour. 
A larger sample of clusters is required to unambiguously address this issue.

Finally, we note that disregarding the division by mass, at all radii the fraction of AGN
in the blue category is twice that in the red. This is broadly consistent with
our earlier results 
that galaxies which show signatures of recent interactions show elevated 
levels of both AGN activity and blue colours (Shabala et al.\ 2012; Kaviraj et al.\ 2012).

%
%
\begin{table*}
\begin{center}
\caption{AGN fractions derived from Fig.~\ref{fig:colmass}, divided by
radius, colour and mass. The colour dividing line for blue cloud and
red sequence is taken as the dashed line from Fig.~\ref{fig:colmass},
whereas the division in mass between massive and low mass galaxies 
is at log(stellar mass)=11.0.
\hfil}
\begin{tabular}{lllll}
\hline
Radius  & Massive        & Low Mass      & Massive        & Low Mass \\
($R_{\mathrm virial}$)   & Red Sequence   & Red Sequence  & Blue Cloud     & Blue Cloud \\
\hline
0--1 & $0.20\pm0.06$ (10/51) & $0.06\pm0.02$ (13/207) & $0.00\pm0.00$ (0/2) & $0.15\pm0.06$ (6/40)\\
1--2 & $0.28\pm0.09$ (10/35) & $0.08\pm0.02$ (10/128) & $1.00\pm1.00$ (1/1) & $0.22\pm0.07$ (11/50)\\
2--3 & $0.19\pm0.11$ (3/16) & $0.05\pm0.02$ (6/120) & $0.00\pm0.00$ (0/4) & $0.23\pm0.07$ (11/47)\\
\hline
\end{tabular}
  \label{tab:colmassfracs}
\end{center}
\end{table*}

\section{Summary}
In summary, we have investigated the AGN fraction
in 6 `clean' galaxy clusters down to $\approx M^{\star} + 1$
as a function of both mass and radius from
the cluster centre. Our main results are:\\
(i) The radial AGN fraction increases steeply in the central
1.5 $R_{\mathrm virial}$ of the composite cluster, but flattens off 
quickly and even decreases beyond this radius. If one were to
compare the central regions of clusters with field samples,
then no difference would be found on the basis of this work.\\
(ii) The AGN fraction by mass shows a significant trend such 
that more massive galaxies are more likely to host AGN.  
Indeed, massive galaxies host more AGN at all radii from the
cluster centre.  The reported deficit of AGN in cluster 
centres may therefore simply be a product of the changing
mix of galaxy types with radius.\\
(iii) Retired AGN (found using the WHAN diagnostic) are found at all radii in the cluster, 
but their response to mass is much more pronounced: we find that massive galaxies 
are more likely to be in the retired class.\\
(iv) AGN have no preferential position
inside galaxy clusters (either with regard to infalling
status, or enhanced local galaxy density).  
This conclusion can be brought in to line with studies of X-ray AGN 
(e.g.\ Haines et al.\ 2012) by considering only the most powerful optical AGN.
These galaxies avoid $|\Delta(cz)|/\sigma_{cz}<1.2$ and may therefore reside on
the cluster caustics (infall regions) as demonstrated in Haines et al.\ (2012).\\
(v) If interactions
with other galaxies trigger AGN activity, then the time-lag
between the trigger and AGN enhancement must be sufficiently
long to mask the site of the encounter and eliminate any signal in 
local galaxy density.

Our favoured scenario for AGN triggering remains a gas-rich interaction,
although increased numbers of galaxies are required to produce better
statistics to firm this speculation up.

\section*{Acknowledgements}
We thank the anonymous referee for his/her interesting feedback
that improved this manuscript.
We further thank David Atlee, 
Heath Jones and Duncan Galloway for stimulating
conversations on this work. 

S.S.S.~thanks the Australian Research Council for the award
of a Super Science Fellowship at the University of Tasmania.
D.J.E.F.~acknowledges support from the Australian Research Council 
via Discovery Project grant DP110102174.

Funding for the SDSS and SDSS-II has been provided by the
Alfred P.\ Sloan Foundation, the Participating Institutions,
the National Science Foundation, the U.S. Department of Energy,
the National Aeronautics and Space Administration, the Japanese
Monbukagakusho, the Max Planck Society, and the Higher Education
Funding Council for England. The SDSS Web Site is http://www.sdss.org/.

The SDSS is managed by the Astrophysical Research Consortium for the
Participating Institutions. The Participating Institutions are the
American Museum of Natural History, Astrophysical Institute Potsdam,
University of Basel, University of Cambridge, Case Western Reserve
University, University of Chicago, Drexel University, Fermilab, the
Institute for Advanced Study, the Japan Participation Group, Johns
Hopkins University, the Joint Institute for Nuclear Astrophysics, the
Kavli Institute for Particle Astrophysics and Cosmology, the Korean
Scientist Group, the Chinese Academy of Sciences (LAMOST), Los Alamos
National Laboratory, the Max-Planck-Institute for Astronomy (MPIA), the
Max-Planck-Institute for Astrophysics (MPA), New Mexico State University,
Ohio State University, University of Pittsburgh, University of Portsmouth,
Princeton University, the United States Naval Observatory, and the University of Washington.

This research has made use of the NASA/IPAC Extragalactic Database (NED) which is
operated by the Jet Propulsion Laboratory, California Institute of Technology,
under contract with the National Aeronautics and Space Administration.

\section*{Appendix~A: Abell~1620}

Abell~1620 is noted in Section~2 as possessing significant substructure
at $>2R_{\mathrm virial}$, in agreement with Burgett et al.\ (2004). 

In Fig.~\ref{fig:a1620} we plot a smoothed surface density of galaxy members
for this cluster.  This figure reveals two overdensities of galaxies 
to the South East of Abell~1620 proper.
We identify these two overdensities as
SDSS-C4 1010 (Miller et al.\ 2005) at
12h48m02.7s -01d39m10s, and
NSC J124857-015532 (Gal et al.\ 2003)
at 12h48m57.7s -01d55m33s using NED.

These sub-clusters (i.e.\ groups) are 
notable in our analysis.  In Fig.~\ref{fig:sigma}, there is a
local peak in $\Sigma_5$ at $1.2R_{\mathrm virial}$--$1.5R_{\mathrm virial}$.  We associate this peak
with these two groups.  
Of note, there are two AGN contained in this peak (i.e.\ with
$\log(\Sigma_5)>1.8$).  It is these AGN that result in the enhanced $\Sigma_5$ average
value noted in Table~\ref{tab:med} at these radii. 
We explicitly note that the mass of these secondary peaks are less than the
primary A1620 peak (as confirmed by Burgett et al.\ 2004).
We further emphasize that removal of 
Abell~1620 from our analysis does not change the primary 
results contained in our work.  
Therefore, we regard these two overdensities to have had 
negligible effect on the AGN contained within it (i.e.\ no
enhancement), in-line with our conclusions.

\end{document}